\documentclass[aps,prd,twocolumns,eqsecnum,preprintnumbers,showpacs,amsmath,amssymb]
{revtex4}

\usepackage{graphicx}
\usepackage{dcolumn}
\usepackage{bm}

\newcommand{\dd}{{\textrm d}}
\newcommand{\GeV}{{\textrm{GeV}}}
\newcommand{\MeV}{{\textrm{MeV}}}

\begin{document}

\title{THE NON-PERTURBATIVE EQUATION OF STATE \\ FOR THE GLUON MATTER}

\author{V. Gogokhia }
\email[]{gogohia@rmki.kfki.hu}

\author{V.V. Skokov }
\email[]{vvskokov@theor.jinr.ru}

\affiliation{HAS, CRIP, RMKI, Depart. Theor. Phys., Budapest 114,
P.O.B. 49, H-1525, Hungary}

\affiliation{BLTP, JINR, 141980, Dubna, Russia}

\date{\today}

\begin{abstract}
In order to derive equation of state for the pure $SU(3)$
Yang-Mills fields from first principles, it is proposed to
generalize the effective potential approach for composite
operators to non-zero temperatures. It is essentially
non-perturbative by construction, since it assumes the summation
of an infinite number of the corresponding contributions. There is
no dependence on the coupling constant, only a dependence on the
mass gap, which is responsible for the large-scale structure of
the QCD ground state. The equation of state generalizes the Bag
constant at non-zero temperatures, while its nontrivial Yang-Mills
part has been approximated by the generalization of the free gluon
propagator to non-zero temperatures, as a first necessary step.
Even in this case we were able to show explicitly that the
pressure may almost continuously change its regime at $T^* = 266.5
\ \MeV$. All the other thermodynamical quantities such as energy
density, entropy, etc. are to be understood to have drastic
changes in their regimes in the close vicinity of $T^*$. All this
is in qualitative and quantitative agreement with thermal lattice
QCD results for the pure Yang-Mills fields. We have firmly
established the behavior of all the thermodynamical quantities in
the region of low temperatures, where thermal lattice QCD
calculations suffer from big uncertainties.
\end{abstract}



\maketitle

\section{Introduction}

The prediction of a possible existence of the Quark-Gluon Plasma
(QGP) was one of the best theoretical achievements of Quantum
Chromodynamics (QCD) at non-zero temperatures and densities (the
rather full list of the corresponding references can be found in
the text-book on finite-temperature field theory in Ref. \cite{1}
and in Ref. \cite{2} as well). The equation of state (EoS) for the
QGP has been derived analytically up to the order $g^6 \ln(1/g^2)$
by using the perturbation theory (PT) expansion for the evaluation
of the corresponding thermodynamical potential term by term
(\cite{1,2,3} and references therein).

However, the most characteristic feature of this PT expansion is
its non-analytical dependence on the QCD coupling constant $g^2$.
In fact, this means that the PT QCD is not applicable at finite
temperatures, apart from maybe at very high temperatures. The
problem is not in poor convergence of this series \cite{1,2,3} (in
mathematics there exist methods how to improve convergence). The
problem is that in the case of the above-mentioned non-analytical
dependence one cannot even define the radius of convergence, so
any next calculated term can be bigger than the previous one. This
is the principle problem which can be resolved by no means. From
the strictly mathematical point of view the four-dimensional QCD
at non-zero temperatures effectively becomes the three-dimensional
theory. At the same time, the three-dimensional QCD has much more
severe infrared singularities \cite{4} and its coupling constant
becomes dimensional. That is the reason why the dependence becomes
not analytical, while using the dimensionless coupling constat
$g^2$ (one needs to introduce three different scales, $T$, $gT$
and $g^2T$, where $T$ is the temperature in order to somehow
understand the dynamics of the QGP within the thermal PT QCD
approach). Thus there is an exact indication that the analytical
EoS derived by thermal PT QCD is wrong.

At present, the only method to be used in order to investigate
thermal QCD is the lattice QCD at finite temperature and baryon
density which underwent a rapid recent progress (\cite{1,2,5,6,7}
and references therein). However, the lattice QCD, being a very
specific regularization scheme, first of all is aimed at obtaining
the well-defined corresponding expressions in order to get correct
numbers from them. So, one gets numbers, but not understanding on
what is going on. Such kind of understanding can only come from
the dynamical theory which is continuous QCD. For example, any
description of the QGP is to be formulated in the framework of the
dynamical theory. The lattice thermal QCD is useless in this. The
need in the analytical EoS remains, but, of course it should be
essentially non-perturbative (NP), reproducing the thermal PT QCD
results at a very high temperature only. Thus analytic NP QCD and
lattice QCD approaches to finite-temperature QCD do not exclude
each other, but contrary they should complement each other.
Especially this is true for low temperatures where lattice QCD
calculations suffer from big uncertainties \cite{1,2,5,6,7}. There
already exist an interesting analytic approaches based on
quasi-particle  and liquid model pictures \cite{8} to analyze
results of $SU(3)$ lattice QCD calculations for the QGP EoS.

The formalism we are going to use in order to generalize it to
non-zero temperature is the effective potential approach for
composite operators \cite{9}. It is essentially NP from the very
beginning, since it is dealing with the expansion of the
corresponding skeleton loop contributions (for more detail
description see section 2 and Ref. \cite{10} as well, where it has
been generalized on quark degrees of freedom, but not using the
confinement-type solution for the quark propagator). The main
purpose of this paper is to derive EoS for the gluon matter by
introducing the temperature dependence into the effective
potential approach in a self-consistent way, in particular by
using the confinement-type solution for the full gluon propagator
(see below).

\section{The VED }

\begin{figure}[t]
\begin{center}
\includegraphics[width=0.8\textwidth]{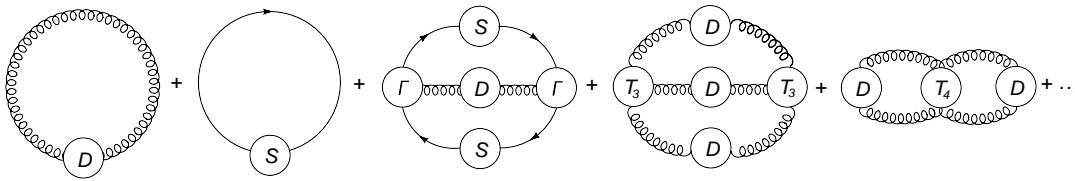}
 \caption{The skeleton loop expansion for the
effective potential. The wavy lines describe the full gluon
propagators $D$, while the solid lines -- the full quark
propagators $S$. $\Gamma$ is the full quark-gluon vertex, while
$T_3$ and $T_4$ are the full three- and four-gluon vertices,
respectively. The ghost skeleton loops are not shown explicitly.}
\label{fig:1}
\end{center}
\end{figure}
\begin{figure}[b]
\begin{center}
\includegraphics[width=0.6\textwidth]{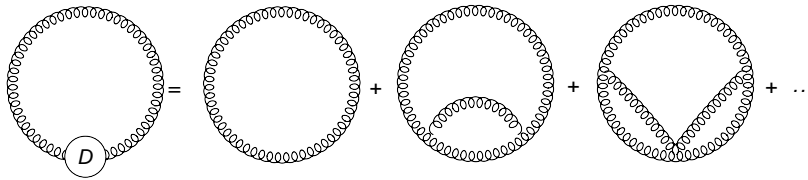}
 \caption{
 Infinite series for the gluon part of the VED (the first skeleton diagram in Fig. 1) }
\label{fig:2}
\end{center}
\end{figure}

The quantum part of the vacuum energy density (VED) is determined
by the effective potential approach for composite operators
\cite{9}. In the absence of external sources the effective
potential is nothing but the VED. It is given in the form of the
skeleton loop expansion, containing all the types of the QCD full
propagators and vertices, see Fig. 1. So each vacuum skeleton loop
itself is a sum of an infinite number of the corresponding PT
vacuum loops, i.e., it contains the point-like vertices and free
propagators (see Fig. 2, where one term only in each lower order
is shown, for simplicity). The number of the vacuum skeleton loops
is equal to the power of the Planck constant, $\hbar$.

Here we are going to formulate a general method of numerical
calculation of the quantum part of the truly NP Yang-Mills (YM)
VED in the covariant gauge QCD. The gluon part of the VED to
leading order (the so-called log-loop level $\sim \hbar$, the
first skeleton loop diagram in Fig. 1, and which PT expansion is
shown explicitly in Fig. 2) is analytically given by the effective
potential for composite operators as follows \cite{9}:

\begin{equation}
V(D) =  { i \over 2} \int {\dd^4q \over (2\pi)^4}
 Tr\{ \ln (D_0^{-1}D) - (D_0^{-1}D) + 1 \},
\end{equation}
where $D(q)$ is the full gluon propagator and $D_0(q)$ is its free
counterpart (see below). Traces over space-time and color group
indices are assumed. Evidently, the effective potential is
normalized to $V(D_0) = 0$. Next-to-leading and higher order
contributions (two and more vacuum skeleton loops) are suppressed
at least by one order of magnitude in powers of $\hbar$. They
reproduce very small numerical corrections to the log-loop terms,
and thus are not important for the numerical calculation of the
VED to leading order.

The two-point Green's function, describing the full gluon
propagator, is

\begin{equation}
D_{\mu\nu}(q) = - i \left\{ T_{\mu\nu}(q)d(-q^2, \xi) + \xi
L_{\mu\nu}(q) \right\} {1 \over q^2 },
\end{equation}
where $\xi$ is the gauge-fixing parameter and

\begin{equation}
T_{\mu\nu}(q) = g_{\mu\nu} - {q_\mu q_\nu \over q^2} = g_{\mu\nu }
- L_{\mu\nu}(q).
\end{equation}

Its free PT counterpart $D_0 \equiv D^0_{\mu\nu}(q)$ is obtained
by putting the full gluon form factor $d(-q^2, \xi)$ in Eq. (2.2)
simply to one, i.e.,

\begin{equation}
D^0_{\mu\nu}(q) = - i \left\{ T_{\mu\nu}(q) + \xi L_{\mu\nu}(q)
\right\} {1 \over q^2}.
\end{equation}

In order to evaluate the effective potential (2.1), on account of
Eqs. (2.2) and (2.4), we use the well-known expression

\begin{equation}
Tr \ln (D_0^{-1}D) = 8 \times 4 \ln det (D_0^{-1}D) = 32 \ln [ (3/
4 )d(-q^2, \xi) + (1 / 4 ) ].
\end{equation}
It becomes zero indeed when equating $d(-q^2, \xi)=1$.

 Going over to four-dimensional Euclidean space in Eq. (2.1),
one obtains ($\epsilon_g = V(D)$)

\begin{equation}
\epsilon_g = - 16 \int {\dd^4q \over (2\pi)^4} \left[ \ln [1 + 3
d(q^2, \xi)] - {3 \over 4}d(q^2, \xi) + a \right],
\end{equation}
where the constant $a = (3/4) - 2 \ln 2 = - 0.6363$ and the
integration from zero to infinity is assumed. The VED $\epsilon_g$
derived in Eq. (2.6) is already a colorless quantity, since it has
been already summed over color indices. Also, only the transversal
("physical") degrees of freedom of gauge bosons contribute to this
equation, so there is no need for ghosts to cancel their
longitudinal (unphysical) counterparts.

However, the derived expression (2.6) remains rather formal, since
it suffers from different types of the PT contributions
("contaminations"). In order to define the truly NP VED free of
all the above-mentioned problems, let us make first the identical
transformation of the full effective charge in Eq. (2.6) as
follows:

\begin{equation}
d(q^2, \xi) = d(q^2, \xi) - d^{PT}(q^2, \xi)  + d^{PT}(q^2, \xi) =
d^{NP}(q^2)  + d^{PT}(q^2, \xi),
\end{equation}
where $d^{PT}(q^2, \xi)$ correctly describes the PT structure of
the full effective charge $d(q^2, \xi)$, including its behavior in
the ultra-violet (UV) limit, compatible with asymptotic freedom
(AF) phenomenon in QCD \cite{11}), otherwise remaining arbitrary.
On the other hand, $d^{NP}(q^2)$ defined by the above-made
subtraction, is assumed to reproduce correctly the NP structure of
the full effective charge, including its asymptotic in the deep
infrared (IR) limit. This underlines the strong intrinsic
influence of the IR properties of the theory on its NP dynamics.
Evidently, both terms are valid in the whole energy/momentum
range, i.e., they are not asymptotics. Let us also emphasize the
principle difference between $d(q^2, \xi)$ and $d^{NP}(q^2)$. The
former is NP quantity "contaminated" by the PT contributions,
while the latter one, being also NP, is, nevertheless, free of
them. Thus the separation between the truly NP effective charge
$d^{NP}(q^2)$ and its nontrivial PT counterpart $d^{PT}(q^2, \xi)$
is achieved. For example, if the full effective charge explicitly
depends on the scale responsible for the truly NP dynamics in QCD,
say $\Lambda^2_{NP}$, then one can define the subtraction

\begin{equation}
d^{NP}(q^2, \Lambda^2_{NP}) = d(q^2, \Lambda^2_{NP}) - d(q^2,
\Lambda^2_{NP} =0) = d(q^2, \Lambda^2_{NP})  - d^{PT}(q^2),
\end{equation}
which is obviously equivalent to the decomposition (2.7). In this
way the above-mentioned separation becomes exact and unique as
well (for such concrete example see below). Let us emphasize that
the dependence of the full effective charge $d(q^2,
\Lambda^2_{NP})$ on $\Lambda^2_{NP}$ can be only regular.
Otherwise it is impossible to assign to it the above-mentioned
physical meaning, since $\Lambda^2_{NP}$ can be only zero (the
formal PT limit) or finite, i.e., it cannot be infinitely large.
In principle, in some special models of the QCD vacuum, for
example such as the Abelian Higgs model \cite{12,13}, the NP scale
is to be identified with the mass of the dual gauge boson. Let us
note that if there is no exact criterion how to distinguish
between the truly NP and the nontrivial PT parts in the full
effective charge as described above, then it is possible from the
full effective charge to subtract its UV asymptotic only.
Evidently, in this case the separation between the truly NP and
the nontrivial PT parts may not be unique.

\section{Generalization to non-zero temperatures}

Substituting the above-discussed exact decomposition (2.7) into
Eq. (2.6), introducing further the effective scale squared,
separating the NP region from the PT one (soft momenta from hard
momenta), and omitting some algebraic rearrangements (see Refs.
\cite{14,15} and especially recent paper \cite{16} for details),
one obtains

\begin{equation}
\epsilon_{YM}(T) = - B_{YM} + B_{YM}(T)+ P_{YM}(T).
\end{equation}
Here evidently $\epsilon_g \equiv \epsilon_{YM}$ and $B_{YM}$ is
the Bag constant at zero temperature \cite{16}. Also, $B_{YM}(T)$
and $P_{YM}(T)$ are explicitly given by the following expressions

\begin{equation}
B_{YM}(T) = 16 \int^{q^2_{eff}} {\dd^4q \over (2\pi)^4} \left[ \ln
[1 + 3 \alpha_s^{NP}(q^2)] - {3 \over 4} \alpha_s^{NP}(q^2)
\right]
\end{equation}
and $P_{YM}(T)$ has more complicate form, namely

\begin{equation}
P_{YM}(T) = - 16  \int {\dd^4q \over (2\pi)^4} \left[ \ln [1 +
3\alpha_s^{PT}(q^2) + 3 \alpha_s^{NP}(q^2)] - {3 \over 4}
[\alpha_s^{PT}(q^2) + \alpha_s^{NP}(q^2)] + a \right],
\end{equation}
respectively, since it depends on both effective charges. In all
these equations

\begin{equation}
\alpha^{NP}_s(q^2) \equiv d^{NP}(q^2), \quad \alpha^{PT}_s(q^2)
\equiv d^{PT}(q^2),
\end{equation}
because $d^{NP}(q^2)$ and $d^{PT}(q^2)$ are the truly NP and the
nontrivial PT effective charges, respectively, as it follows from
above. Precisely these expressions should be generalized to
non-zero temperatures in order to get EoS for the pure YM fields.
That is why we introduce the dependence on the temperature $T$ in
advance. Evidently, Eq. (3.2) will reproduce the
temperature-dependent Bag constant. In the expression for
$P_{YM}(T)$ the integration is from zero to infinity, while in the
integral for $B_{YM}(T)$ it is from zero to the effective scale
squared $q^2_{eff}$, which just symbolically shown in Eq. (3.2).
It is worth emphasizing that a so defined Bag constant (3.2) is
free of all types of PT contributions ("contaminations"), as it is
required (this was a reason for the above-mentioned algebraic
rearrangements and subtractions, see Ref. \cite{16} and references
therein).

The problem remaining to solve is to choose the truly NP effective
charge $\alpha_s^{NP}(q^2)$. For the different truly NP effective
charges we will get different analytical and numerical results.
That is why the choice for its explicit expression should be
physically and mathematically well justified. Let us choose the
truly NP effective charge as follows:

\begin{equation}
\alpha_s^{NP}(q^2) = {\Lambda^2_{NP} \over q^2},
\end{equation}
where $\Lambda_{NP}$ is the mass scale parameter (the mass gap)
responsible for the large-scale structure of the true QCD vacuum.
It is well known that in continuous QCD it leads to the linear
rising potential between heavy quarks, "seen" by lattice QCD
\cite{17,18} as well ($(q^2)^{-2}$-type behavior for the full
gluon propagator). Moreover, in Ref. \cite{19} it has been
explicitly shown that it is a direct nonlinear iteration solution
of the transcendental equation for the full gluon propagator in
the presence of a renormalized mass gap. The separation between
the truly NP and the nontrivial PT effective charges is both exact
and unique, since the PT effective charge is always regular at
zero, while the truly NP effective is singular at the origin (in
the formal PT limit ($\Lambda^2_{NP} \rightarrow 0$) the truly NP
effective charge vanishes, while its nontrivial PT counterpart
will survive). Let us also note that the chosen effective charge
(3.5) does not depend explicitly on the gauge choice. It has been
already used \cite{14,15,16} in order to calculate the Bag
constant, which turned out to be in a very good agreement with
such important phenomenological parameter as the gluon condensate.
It leads to many other desirable properties for the Bag pressure
at zero temperature \cite{16}. Thus, our choice (3.5) is
physically justified and mathematically confirmed, as required
above.

In the imaginary time formalism \cite{1,20}, these expressions can
be easily generalized to non-zero temperatures $T$ according to
the prescription (let us remind that there is already Euclidean
signature)

\begin{equation}
\int {\dd q_0 \over (2\pi)} \rightarrow T \sum_{n=- \infty}^{+
\infty}, \quad \quad q^2 = {\bf q}^2 + q^2_0 = {\bf q}^2 +
\omega^2_n = \omega^2 + \omega^2_n, \quad \omega_n = 2n \pi T,
\end{equation}
i.e., each integral over $q_0$ of the loop momentum is to be
replaced by the sum over Matsubara frequencies labelled by $n$,
which obviously assumes the replacement $q_0 \rightarrow \omega_n=
2n \pi T$ for bosons (gluons). In frequency-momentum space the
truly NP effective charge becomes

\begin{equation}
\alpha_s^{NP}(q^2) = \alpha^{NP}_s({\bf q}^2, \omega_n^2) = {
\Lambda^2_{NP} \over {\bf q}^2 + \omega_n^2} = { \Lambda^2_{NP}
\over \omega^2 + \omega_n^2}, \quad \alpha_s^{PT}(q^2) =
\alpha^{PT}_s({\bf q}^2, \omega_n^2) = \alpha_s^{PT} (\omega^2,
\omega_n^2).
\end{equation}
It is also convenient to introduce the following notations

\begin{equation}
T^{-1} = \beta, \quad \omega = \sqrt{{\bf q}^2},
\end{equation}
where, evidently, in all expressions here and below ${\bf q}^2$ is
the three-dimensional loop momentum squared in complete agreement
with the relations (3.6).

\section{The derivation of $B_{YM}(T)$}

In frequency-momentum space the Bag pressure (3.2) after the
substitution of the relations (3.6) becomes

\begin{equation}
B_{YM}(T) =  16 \int {\dd^3q \over (2\pi)^3} \ T \sum_{n= -
\infty}^{+ \infty} \left[ \ln [1 + 3 \alpha_s^{NP}({\bf q}^2,
\omega^2_n)] - {3 \over 4} \alpha_s^{NP}({\bf q}^2, \omega^2_n)
\right],
\end{equation}
where the truly NP effective charge is given in Eq. (3.7), and
notations of Eqs. (3.7)-(3.8) are also valid, of course. After its
substitution into Eq. (4.1), one yields

\begin{equation}
B_{YM}(T) = 16 \int {\dd^3q \over (2\pi)^3} \ T \sum_{n= -
\infty}^{+ \infty} \left[ \ln [ 3 \Lambda^2_{NP} + {\bf q}^2 +
\omega^2_n] - \ln [{\bf q}^2 + \omega^2_n] - {3 \over 4}
\Lambda^2_{NP} { 1 \over {\bf q}^2 + \omega^2_n} \right].
\end{equation}
The summation over the Matsubara frequencies squared $\omega^2_n=
(2 \pi T)^2 n^2$ can be easily done, and the dependence on the
effective scale $\omega_{eff}$ (see Appendix) is omitted, for
simplicity. Here it is also convenient to introduce the following
notation

\begin{equation}
\omega' = \sqrt{{\bf q}^2 + m'^2_{eff}}= \sqrt{{\bf q}^2 + 3
\Lambda^2_{NP}} = \sqrt{\omega^2 + 3 \Lambda^2_{NP}} = \omega
\sqrt{1 + { 3 \over \omega^2} \Lambda^2_{NP}},
\end{equation}
So it is possible to say that within our approach to non-zero
temperatures we have two sorts of gluons: massless $\omega$ and
massive $\omega'$ with the effective mass

\begin{equation}
m'_{eff}= \sqrt{3} \Lambda_{NP}.
\end{equation}

In the second term the summation over Matsubara frequencies can be
done explicitly, namely

\begin{eqnarray}
\sum_{n= - \infty}^{+ \infty} { 1 \over {\bf q}^2 + \omega^2_n}
&=& \sum_{n= - \infty}^{\infty} { 1 \over \omega^2 + (2\pi T)^2
n^2} = (2\pi / \beta)^{-2} \sum_{n= -
\infty}^{ + \infty} { 1 \over n^2 + (\beta \omega / 2 \pi)^2} \nonumber\\
&=& (2\pi / \beta)^{-2} (2 \pi^2 / \beta \omega) \left( 1 + { 2
\over e^{\beta\omega} -1} \right)=  { \beta \over 2 \omega} \left(
1 + { 2 \over e^{\beta\omega} -1} \right).
\end{eqnarray}

\subsection{The summation of logarithms}

In terms of the above-introduced parameters the sums in Eq. (4.2),
containing the corresponding logarithms, look like

\begin{equation}
\sum_{n= - \infty}^{+ \infty} \ln [ 3 \Lambda^2_{NP} + {\bf q}^2 +
\omega^2_n] = \ln \omega'^2 + 2 \sum_{n= 1}^{ \infty} \ln (2\pi /
\beta)^2[ n^2 + (\beta \omega' / 2\pi)^2]
\end{equation}
and

\begin{equation}
\sum_{n= - \infty}^{+ \infty} \ln [{\bf q}^2 + \omega^2_n] = \ln
\omega^2 + 2 \sum_{n= 1}^{ \infty} \ln (2\pi / \beta)^2[ n^2 +
(\beta \omega / 2\pi)^2].
\end{equation}

It is convenient to introduce the notations as follows:

\begin{equation}
L(\omega') = \sum_{n=1}^{ \infty} \ln[ n^2 + (\beta \omega' /
2\pi)^2] = \sum_{n=1}^{ \infty} \ln n^2  + \sum_{n= 1}^{ \infty}
\ln \left[ 1 - {x'^2 \over n^2 \pi^2} \right]
\end{equation}
and equivalently

\begin{equation}
L(\omega) = \sum_{n=1}^{ \infty} \ln[ n^2 + (\beta \omega /
2\pi)^2] = \sum_{n=1}^{ \infty} \ln n^2  + \sum_{n= 1}^{ \infty}
\ln \left[ 1 - { x^2 \over n^2 \pi^2} \right].
\end{equation}
Calculating explicitly the first one, we can calculate
automatically the second by simply replacing $\omega' \rightarrow
\omega$ and vice-versa. Evidently, in these expressions we
introduce the following notations:

\begin{equation}
x'^2  = - \left( {\beta \omega' \over 2 } \right)^2, \quad x^2 = -
\left( {\beta \omega \over 2 } \right)^2.
\end{equation}
So the difference $L(\omega') - L(\omega)$ becomes

\begin{equation}
L(\omega') - L(\omega) = \sum_{n= 1}^{ \infty} \ln \left[ 1 -
{x'^2 \over n^2 \pi^2} \right] - \sum_{n= 1}^{ \infty} \ln \left[
1 - { x^2 \over n^2 \pi^2} \right] = \ln \sin x' - {1 \over 2} \ln
x'^2 - \ln \sin x + {1 \over 2} \ln x^2,
\end{equation}
or equivalently

\begin{equation}
L(\omega') - L(\omega) = - {1 \over 2} \ln \left( { x'^2 \over
x^2} \right) + \ln \left( {\sin x'  \over \sin x} \right).
\end{equation}
From the relation (4.10) it follows

\begin{equation}
x'  = \pm i \left( {\beta \omega' \over 2 } \right), \quad x = \pm
i \left( {\beta \omega \over 2 } \right),
\end{equation}
so the previous equation (4.12) finally becomes

\begin{equation}
 L(\omega') - L(\omega) = - { 1 \over 2} \ln \left( {
\omega'^2 \over \omega^2} \right) + { 1 \over 2} \beta( \omega' -
\omega) + \ln \left({1 - e^{- \beta \omega'} \over 1 - e^{- \beta
\omega}} \right).
\end{equation}

\subsection{The explicit expressions for the integrals}

Substituting all our results of the summations into Eq. (4.2),
dropping a $\beta$-independent terms \cite{1}, and performing
almost trivial integration over angular variables, one obtains

\begin{equation}
B_{YM}(T) =  - {8 \over \pi^2} \int \dd\omega \ \omega^2 \left[ {3
\over 4} \Lambda^2_{NP} {1 \over \omega} { 1 \over e^{\beta
\omega} - 1 } - 2 \beta^{-1} \ln \left( {1 - e^{- \beta \omega'}
\over 1 - e^{- \beta \omega}} \right) \right].
\end{equation}

It is convenient to present the integral (4.15) as a sum of a few
terms

\begin{equation}
B_{YM}(T) = - {6 \over \pi^2} \Lambda^2_{NP} B_{YM}^{(1)}(T) - {16
\over \pi^2} T \left[ B_{YM}^{(2)}(T) - B_{YM}^{(3)}(T) \right],
\end{equation}
where the explicit expressions of all these integrals are given
below

\begin{equation}
B_{YM}^{(1)}(T) = \int_0^{\omega_{eff}} \dd\omega {\omega \over
e^{\beta\omega} -1},
\end{equation}

\begin{equation}
B_{YM}^{(2)}(T)  = \int_0^{\omega_{eff}} \dd\omega \ \omega^2 \ln
\left( 1- e^{-\beta \omega} \right),
\end{equation}

\begin{equation}
B_{YM}^{(3)}(T)  = \int_0^{\omega_{eff}} \dd\omega \ \omega^2 \ln
\left( 1- e^{-\beta \omega'} \right).
\end{equation}
In all these integrals the upper limit $\omega_{eff}$ is
explicitly shown now and $\beta^{-1} =T$, while

\begin{equation}
\omega' = \sqrt{ \omega^2 + 3 \Lambda^2_{NP}} = \omega \sqrt{1 + {
3 \Lambda^2_{NP} \over \omega^2}}.
\end{equation}

\section{The derivation of $P_{YM}(T)$}

The term which contains the information about the nontrivial YM
part (3.3) of the future gluon plasma EoS is

\begin{equation}
P_{YM}(T) = - 16  \int {\dd^4q \over (2\pi)^4} \left[ \ln [1 +
3\alpha_s^{PT}(q^2) + 3 \alpha_s^{NP}(q^2)] - {3 \over 4}
[\alpha_s^{PT}(q^2) + \alpha_s^{NP}(q^2)] + a \right],
\end{equation}
where $\alpha_s^{PT}(q^2)$ is the nontrivial PT effective charge.
Due to the above-mentioned normalization of the effective
potential approach in Eq. (2.1), the investigation of this part
makes sense to begin with the approximation of the nontrivial PT
part by its free PT counterpart, i.e., to put
$\alpha_s^{PT}(q^2)=1$, as a first necessary step. Then Eq. (5.1)
for the YM pressure in frequency-momentum space becomes

\begin{equation}
P_{YM}(T) = - 16 \int {\dd^3q \over (2\pi)^3} \ T \sum_{n= -
\infty}^{+ \infty} \left[ \ln [1 + {3 \over 4} \alpha_s^{NP}({\bf
q}^2, \omega^2_n)]  - {3 \over 4} \alpha_s^{NP} ({\bf q}^2,
\omega^2_n) \right],
\end{equation}
and after substituting of the relations (3.7) into it, one obtains

\begin{equation}
P_{YM}(T) = - 16 \int {\dd^3q \over (2\pi)^3} \ T \sum_{n= -
\infty}^{+ \infty} \left[ \ln [ {3 \over 4} \Lambda^2_{NP} + {\bf
q}^2 + \omega^2_n] - \ln [{\bf q}^2 + \omega^2_n] - {3 \over 4}
\Lambda^2_{NP} { 1 \over {\bf q}^2 + \omega^2_n} \right],
\end{equation}
and the summation over the Matsubara frequencies squared
$\omega^2_n= (2 \pi T)^2 n^2$ can be easily done. For this purpose
it is also convenient to introduce the following notation

\begin{equation}
\bar \omega = \sqrt{{\bf q}^2 + \bar m^2_{eff}}= \sqrt{{\bf q}^2 +
{3 \over 4} \Lambda^2_{NP}} = \sqrt{\omega^2 + {3 \over 4}
\Lambda^2_{NP}} = \omega \sqrt{1 + { 3 \over 4 \omega^2}
\Lambda^2_{NP}},
\end{equation}
so it is possible to say that within our approach to non-zero
temperatures at this intermediate stage we have two sorts of
gluons: massless $\omega$ and massive $\bar \omega$ with the
effective mass

\begin{equation}
\bar m_{eff}= {\sqrt{3} \over 2} \Lambda_{NP} = {1 \over 2}
m'_{eff}
\end{equation}

Comparing Eqs.(4.2) and (5.3) one can write down the final result
directly. For this purpose, in the final system of Eqs.
(4.16)-(4.19) one must change the overall sign, replace $\omega'$
by $\bar \omega$ and integrate from zero to infinity. Thus, one
obtains

\begin{equation}
P_{YM}(T) = {6 \over \pi^2} \Lambda^2_{NP} P_{YM}^{(1)}(T) + {16
\over \pi^2} T \left[ P_{YM}^{(2)}(T) - P_{YM}^{(3)}(T) \right],
\end{equation}
where the explicit expressions of all these integrals are given
below

\begin{equation}
P_{YM}^{(1)}(T) = \int_0^{\infty} \dd\omega {\omega \over
e^{\beta\omega} -1},
\end{equation}

\begin{equation}
P_{YM}^{(2)}(T)  = \int_0^{\infty} \dd\omega \ \omega^2 \ln \left(
1- e^{-\beta \omega} \right),
\end{equation}

\begin{equation}
P_{YM}^{(3)}(T)  = \int_0^{\infty} \dd\omega \ \omega^2 \ln \left(
1- e^{-\beta \bar \omega} \right).
\end{equation}

\section{The gluon matter EoS}

Denoting further $\epsilon_{YM}(T) + B_{YM} = P_{GM}(T)$ in the
left-hand-side of our EoS (3.1), one obtains

\begin{equation}
P_{GM}(T) =  B_{YM}(T) + P_{YM}(T),
\end{equation}
and in this equation $B_{YM}(T)$ and $P_{YM}(T)$ are given in Eqs.
(4.16) and (5.6), respectively. Summing up all the integrals
(4.17)-(4.19) and (5.7)-(5.9), one obtains that the gluon matter
EoS (6.1) finally becomes,

\begin{equation}
P_{GM}(T) = {6 \over \pi^2} \Lambda^2_{NP} P_1 (T) + {16 \over
\pi^2} T [P_2(T) + P_3(T) - P_4(T)],
\end{equation}
where the dependence on the thermodynamical variable $T$ is only
shown explicitly and

\begin{equation}
P_1(T) = \int_{\omega_{eff}}^{\infty} \dd \omega {\omega \over
e^{\beta\omega} -1},
\end{equation}
while

\begin{eqnarray}
P_2(T) &=& \int_{\omega_{eff}}^{\infty} \dd \omega \ \omega^2
\ln \left( 1- e^{-\beta\omega} \right), \nonumber\\
P_3(T)&=& \int_0^{\omega_{eff}} \dd \omega \ \omega^2 \ln
\left( 1 - e^{- \beta\omega'} \right), \nonumber\\
P_4(T) &=& \int_0^{\infty} \dd \omega \ \omega^2 \ln \left( 1 -
e^{- \beta \bar \omega} \right).
\end{eqnarray}
Let us recall once more that in all integrals $\beta = T^{-1}$ and
$\omega_{eff}$ is fixed (see Appendix), while

\begin{equation}
\bar \omega = \sqrt{ \omega^2 + {3 \over 4} \Lambda^2_{NP}}, \quad
\omega' = \sqrt{ \omega^2 + 3 \Lambda^2_{NP}}.
\end{equation}
In the formal PT limit ($\Lambda^2_{NP}=0$) from these relations
it follows that $\bar \omega = \omega' = \omega$ and the
combination $P_2(T) + P_3(T) - P_4(T)$ becomes identical zero.
Thus the gluon matter pressure (6.2) in this limit vanishes, i.e.,
it is truly NP, indeed.

The effective potential has been normalized to zero in the $D
\rightarrow D_0$ limit, which reproduces the so-called
Stefan-Boltzmann (SB) non-interacting (ideal) gas of massless
particles (gluons) at high temperatures \cite{1}. So the SB limit
can be added (if necessary) to the truly NP pressure (6.2) in the
$T \rightarrow \infty$ limit only, i.e.,

\begin{equation}
P_{GM}(T) \rightarrow P_{SB}=  { 8 \over 45} \pi^2 T^4, \quad T
\rightarrow \infty \ (\beta \rightarrow 0).
\end{equation}
In the same way, the corresponding SB limits should be added (if
necessary) to all other thermodynamical quantities considered
below.

\section{Thermodynamical potential and other thermodynamical quantities}

In quantum statistics the thermodynamical potential $\Omega(T)$ is
nothing but the pressure $P(T)$ apart from the sign, i,e, in our
case we can put

\begin{equation}
\Omega(T) = - P_{GM}(T).
\end{equation}
In quantum statistical theory all the important quantities such as
energy density, entropy, etc., are to be expressed in terms of the
thermodynamical potential. However, in the truly NP approach we
cannot use the trivial relations between them which traced back to
the PT even at non-zero temperatures. So the general formulae
which to be used are \cite{1}

\begin{eqnarray}
\epsilon(T) &=& - T \left( \partial \Omega(T) \over \partial T
\right) +
\Omega, \nonumber\\
s(T) &=& -  {\partial \Omega(T) \over \partial T}
\end{eqnarray}
for the pure YM fields, i.e., when the chemical potential is equal
to zero. Evidently, here and everywhere below $\epsilon$ and $s$
are energy density and entropy, respectively, of the pure NP gluon
matter.

\subsection{The energy density}

From Eqs. (7.1)-(7.2) it follows that

\begin{equation}
\epsilon(T) = T s(T) - P(T),
\end{equation}
so substituting the corresponding explicit expressions (7.1) and
(6.2) and doing some algebra, one obtains

\begin{equation}
\epsilon(T) = {6 \over \pi^2} \Lambda^2_{NP} \left( T {\dd P_1(T)
\over \dd T} - P_1(T) \right) + {16 \over \pi^2} T^2 {\dd \over
\dd T} M(T),
\end{equation}
where obviously $\dd \equiv \partial$, since the only dependence
on $T$ is present. Here and everywhere below we introduced the
following notation:

\begin{equation}
M(T) =  P_2(T) + P_3(T) - P_4(T).
\end{equation}
Also, here and below all the integrals and their derivatives can
be explicitly obtained from the expressions (6.3)-(6.5). Again the
SB energy density

\begin{equation}
\epsilon_{SB}(T) = { 24 \over 45} \pi^2 T^4
\end{equation}
should be added to our expression (7.4) in the high temperature $T
\rightarrow \infty \ (\beta \rightarrow 0)$ limit only.

\subsection{The entropy}

In the same way the entropy (7.2) becomes

\begin{equation}
s(T) =  {6 \over \pi^2} \Lambda^2_{NP} {\dd P_1(T) \over \dd T} +
{16 \over \pi^2} M(T) + {16 \over \pi^2} T {\dd \over \dd T} M(T),
\end{equation}
and again the SB entropy

\begin{equation}
s_{SB}(T) =  { 32 \over 45} \pi^2 T^3.
\end{equation}
should be added to our expression (7.7) in the high temperature $T
\rightarrow \infty \ (\beta \rightarrow 0)$ limit only.

\subsection{The heat capacity}

One of the interesting thermodynamical characteristics of the QGP
is the heat capacity $c_V$, which is defined as the derivative of
the energy density. Then from the thermodynamical relations
(7.1)-(7.2) it follows

\begin{equation}
c_V(T) = { \partial \epsilon(T) \over \partial T}  = T \left(
\partial s(T) \over \partial T \right).
\end{equation}

Using the explicit expression for the energy density (7.4), one
finally obtains

\begin{equation}
c_V(T) = {6 \over \pi^2} \Lambda^2_{NP} T {\dd^2 P_1(T) \over \dd
T^2} + {32 \over \pi^2} T {\dd \over \dd T}M(T) + {16 \over \pi^2}
T^2 {\dd^2 \over \dd T^2}M(T).
\end{equation}
As in previous cases, the SB heat capacity

\begin{equation}
c_V^{SB}(T) = { 96 \over 45} \pi^2 T^3
\end{equation}
should be added to our expression (7.10) in the high temperature
$T \rightarrow \infty \ (\beta \rightarrow 0)$ limit only.

\begin{figure}
\begin{center}
\includegraphics[width=9cm]{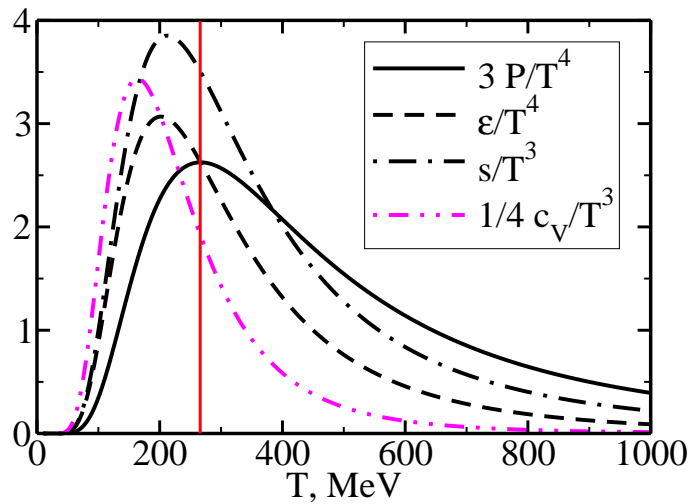}
\caption{The NP pressure $P$, energy density $\epsilon$, entropy
$s$ and heat capacity $c_{V}$ as a functions of the temperature.
The NP gluon pressure $P$ has a maximum at $T^*=266.5 \ \MeV$.}
\label{fig:3}
\end{center}
\end{figure}

\section{Numerical results and discussion}

All our numerical results are present in Fig. 3. It is seen
explicitly that the NP gluon pressure may almost continuously
change its regime in the close neighborhood of a maximum at $T^* =
266.5 \ \MeV$ in order to achieve the thermodynamical SB limit at
high temperatures. For the displayed quantities in Fig. 3 the SB
limits are the corresponding constants. At the same time, for all
other thermodynamical quantities such as the energy density,
entropy and heat capacity this is impossible as it follows from
the curves shown in Fig. 3 (none of their power-type fall off at
this point can be smoothly transformed into the constant behavior
at high temperatures). In order to achieve the thermodynamical SB
limits at high temperatures their full counter-parts should
undergo drastic changes in their regimes in the close neighborhood
of this point. As we already know from thermodynamics of $SU(3)$
lattice QCD \cite{1,2,21} the energy and entropy densities have a
discontinuity at a point $T_c=260 \ \MeV$, while the pressure
remains continuous. Our characteristic temperature $T^*=266.5 \
\MeV$ is, surprisingly, very close to the same value. A clear
evidence that something nontrivial in the behavior of the
thermodynamical quantities in the vicinity of our characteristic
temperature $T^* = 266.5 \ \MeV$ should actually take place
follows from the fact that at this point $\epsilon = 3 P$, which
should be valid at a very high temperatures only (SB limit). In
other words, in order to derive EoS valid above $T^*$, and thus to
provide a correct picture of thermodynamics of the gluon matter in
the whole range of temperature, one needs the nontrivial
approximation of the YM part (5.1), compatible with AF phenomenon
in QCD \cite{11}. Just this will be subject of the subsequent
paper.

If we were not aware of the thermal lattice QCD results then we
would be able to predict them. But we aware of them, so lattice
results confirm our expectations of a sharp changes in the
behavior of the entropy and energy densities in the region where
the pressure is continuous. At the same time, it is worth
emphasizing that we have no any problems in describing the
behavior of all the important thermodynamical quantities at low
temperatures below $T^*$ (see Fig. 3). Moreover, apparently for
the first time it is possible to predict their behavior in the
region of low temperatures within our approach (there are no
convincing lattice data for this region). We do not expect any
serious changes in the behavior of the thermodynamical quantities
in this region (exponential fall off or rise when the temperature
goes down or up, respectively) even after taking into account the
above-mentioned nontrivial approximation of the YM part (5.1),
apart from "non-physical" maximums which should disappear, of
course. However, whatever changes may occur they will be under our
control.

The confinement dynamics (3.5) generalized on zero-temperatures in
Eq. (3.7) is still important especially in the region of low
temperatures even up to the temperature at which all the important
thermodynamical quantities may undergo drastic changes in their
behavior (apart from the pressure). From the structure of our EoS
(see Eqs. (6.2)-(6.5)), it clearly follows that below
$\omega_{eff}$ (which fixes $T^*$, there is no explicit dependence
between them, but rather a correspondence) and thus below $T^*$ we
have mainly the massive gluon excitations $\omega'$ of the
dynamical origin, which can be interpreted as the glueballs with
masses $m'_{eff}=\sqrt{3} \Lambda_{NP} = 1.17 \ \GeV$. Above $T^*$
the gluon matter consists mainly of the free gluons $\omega$
($\bar \omega$ gluons are artifacts due to the approximation of
the nontrivial PT effective charge by its free PT counterpart in
Eq. (5.1), as well as the above-mentioned "non-physical" maximums
and hence their "tails" at high temperatures in Fig. 3). Just the
confinement dynamics determines the phase transition from
glueballs to "free" gluons and vice-versa in the case of $SU(3)$
YM fields within our approach, indeed.

\begin{acknowledgments}

Support from HAS-JINR Agreement, RFBR grant No. 05-02-17695 and
grant RNP 2.1.1.5409 is to be acknowledged. We would like to thank
P. Levai for useful discussions and remarks. We also grateful to
J. Nyiri for help.

\end{acknowledgments}

\appendix
\section{The scale-setting scheme}

From the relations (3.6) it follows that in frequency-momentum
space a possible free parameter of our approach is the effective
scale

\begin{equation}
\omega_{eff} = \sqrt{q^2_{eff}  - \omega^2_c},
\end{equation}
where we introduced the constant Matsubara frequency $\omega_c$,
which is always positive. So $\omega_{eff}$ is always less or
equal to $q_{eff}$ of the four-dimensional QCD, i.e.,

\begin{equation}
\omega_{eff} \leq q_{eff}.
\end{equation}
One then can conclude that $q_{eff}$ is a very good upper limit
for $\omega_{eff}$. In this connection, let us recall now that the
Bag constant $B_{YM}$ at zero temperatures has been successfully
calculated at a scale $q^2_{eff} = 1 \ \GeV^2$, in fair agreement
with other phenomenological quantities such as gluon condensate
\cite{16}. So let us fixed the effective scale $\omega_{eff}$ as
follows:

\begin{equation}
\omega_{eff} = q_{eff} = 1 \ \GeV.
\end{equation}

The mass gap squared $\Lambda^2_{NP}$ calculated just at this
scale is equal to \cite{16}

\begin{equation}
\Lambda^2_{NP} = 0.4564 \ \GeV^2.
\end{equation}

Thus, we have no free parameters in our approach. The confinement
dynamics is nontrivially taken into account directly through the
mass gap, and not through the Bag constant itself.


\begin{thebibliography}{}
\bibitem{1}
   J.I. Kapusta, C. Gale, Finite-Temperature Field Theory
   (Cambridge University Press, 2006).
\bibitem{2}
   Quark Matter 2005, Edited by T. Csorgo, G. David, P. Levai, G.
   Papp (ELSEVIER, Amsterdam-...-St. Louis, 2005); \\
   M. Gyulassy, L. McLerran, arXiv:nucl-th/0405013.
\bibitem{3}
   K. Kajantie, M. Lane, K. Rummukainen, Y. Schroder, Phys. Rev. D (67) (2003) 105008.
\bibitem{4}
   I.M. Gelfand, G.E. Shilov, Generalized Functions, v. I, (AP,
   1964).
\bibitem{5}
    Y.~Aoki, Z.~Fodor, S.~D.~Katz and K.~K.~Szabo,
    JHEP {\bf 0601} (2006) 089
    [arXiv:hep-lat/0510084].
\bibitem{6}
    C.~Schmidt, Z.~Fodor and S.~D.~Katz,
  PoS {\bf LAT2005} (2006) 163
  [arXiv:hep-lat/0510087].
\bibitem{7}
  M.~Cheng {\it et al.},
  arXiv:0710.0354 [hep-lat].
\bibitem{8}
   P. Levai, U. Heinz,
     Phys.\ Rev.\  C {\bf 57}, (1998) 1879
   [arXiv:hep-ph/9710463]; \\
  A.~Peshier, B.~Kampfer, O.~P.~Pavlenko and G.~Soff,
  Phys.\ Rev.\  D {\bf 54}, (1996) 2399;\\
  A.~Peshier, B.~Kampfer and G.~Soff,
  Phys.\ Rev.\  C {\bf 61}, (2000) 045203
  [arXiv:hep-ph/9911474];\\
     K.~K.~Szabo and A.~I.~Toth,
  JHEP {\bf 0306}, (2003) 008
  [arXiv:hep-ph/0302255];\\
      M.~A.~Thaler, R.~A.~Schneider and W.~Weise,
  Phys.\ Rev.\  C {\bf 69}, (2004) 035210;\\
    Yu.~B.~Ivanov, V.~V.~Skokov and V.~D.~Toneev,
   Phys.\ Rev.\  D {\bf 71} (2005) 014005
  [arXiv:hep-ph/0410127];\\
    C.~Ratti, S.~Roessner, M.~A.~Thaler and W.~Weise,
  Eur.\ Phys.\ J.\  C {\bf 49} (2007) 213
  [arXiv:hep-ph/0609218];\\
    M.~Bluhm, B.~Kampfer and G.~Soff,
  Phys.\ Lett.\  B {\bf 620} (2005) 131
  [arXiv:hep-ph/0411106];\\
    W.~Cassing, Nucl.\ Phys.\  A {\bf 791} (2007) 365 [arXiv:0704.1410
    [nucl-th]].\\
    J. Letessier, J. Rafelski, arXiv:hep-ph/0301099.
\bibitem{9}
   J.M. Cornwall, R. Jackiw, E. Tomboulis, Phys. Rev. D 10 (1974) 2428.
\bibitem{10}
   T. Schafer, Nucl. Phys. B 575 (2000) 269.
\bibitem{11}
   W. Marciano, H. Pagels, Phys. Rep. C 36 (1978) 137.
\bibitem{12}
   M. Baker, J.S. Ball, F. Zachariasen, Phys. Rev. D 37 (1988) 1036.
\bibitem{13}
   M.N. Chernodub, M.I. Polikarpov, V.I. Zakharov, hep-ph/9903272.
\bibitem{14}
   V. Gogohia, Gy. Kluge, Phys. Rev. D 62 (2000) 076008
\bibitem{15}
   V. Gogohia, H. Toki, T. Sakai, Gy. Kluge, Int. Jour. Mod. Phys.
   A 15 (2000) 45.
\bibitem{16}
   V. Gogokhia, G.G. Barnafoldi, arXiv:0708.0163v2 [hep-ph];\\
   V. Gogokhia, arXiv:hep-ph/0508224.
\bibitem{17}
   K.D. Born et al., Phys. Lett. B 329 (1994) 325.
\bibitem{18}
   V.M. Miller et al., Phys. Lett. B 335 (1994) 71.
\bibitem{19}
   V. Gogokhia, hep-ph/0702066.
\bibitem{20}
   L. Dolan, R. Jakiw, Phys. Rev. D 9 (1974) 3320.
\bibitem{21}
  G.~Boyd, J.~Engels, F.~Karsch, E.~Laermann, C.~Legeland, M.~Lutgemeier and B.~Petersson,
  Nucl. Phys. B 469 (1996) 419, \
  [arXiv:hep-lat/9602007].
\end{thebibliography}
\end{document}